**Assessment of climate change effects on mountain ecosystems through a cross-site analysis in the Alps and Apennines.**

Rogora M.[1*a], Frate L.[2a], Carranza M.L.[2a], Freppaz M.[3a], Stanisci A.[2a], Bertani I.[4], Bottarin R.[5], Brambilla A.[6], Canullo R.[7], Carbognani M.[8], Cerrato C.[9], Chelli S.[7], Cremonese E.[10], Cutini M.[11], Di Musciano M.[12], Erschbamer B.[13], Godone D.[14], Iocchi M.[11], Isabellon M.[3,10], Magnani A.[15], Mazzola L.[16], Morra di Cella U.[17], Pauli H.[18], Petey M.[17], Petriccione B.[19], Porro F.[20], Psenner R.[5,21], Rossetti G.[22], Scotti A.[5], Sommaruga R.[21], Tappeiner U.[5], Theurillat J.-P.[23], Tomaselli M.[8], Viglietti D.[3], Viterbi R.[9], Vittoz P.[24], Winkler M.[18] Matteucci G.[25a]

*Corresponding author: e-mail address: m.rogora@ise.cnr.it (M. Rogora)

[a] joint first authors

[1] CNR Institute of Ecosystem Study, Verbania Pallanza, Italy

[2] DIBT, Envix-Lab, University of Molise, Pesche (IS), Italy

[3] DISAFA, NatRisk, University of Turin, Grugliasco (TO), Turin, Italy

[4] Graham Sustainability Institute, University of Michigan, 625 E. Liberty St., Ann Arbor, MI 48104, USA

[5] Eurac Research, Institute for Alpine Environment, Bolzano (BZ), Italy

[6] Alpine Wilidlife Research Centre, Gran Paradiso National Park, Degioz (AO) 11 Valsavarenche, Italy; Department of Evolutionary Biology and Environmental Studies, University of Zurich. Winterthurerstrasse 190, 8057 Zurich, Switzerland

[7] School of Biosciences and Veterinary Medicine, Plant Diversity and Ecosystems Management Unit, University of Camerino (MC) Italy

[8] Department of Chemistry, Life Sciences and Environmental Sustainability University of Parma, Parma, Italy

[9] Alpine Wilidlife Research Centre, Gran Paradiso National Park, Degioz (AO) 11 Valsavarenche, Italy

[10] Environmental Protection Agency of Aosta Valley, ARPA VdA, Climate Change Unit, Aosta, Italy

[11] Department of Biology, University of Roma Tre, Viale G. Marconi, 446-00146 Rome, Italy

[12] Department of Life Health & Environmental Sciences, University of L'Aquila Via Vetoio, 67100 L'Aquila, Italy

[13] University of Innsbruck, Institute of Botany. Sternwartestr 15, A- 6020 Insbruck, Austria

[14] CNR IRPI Geohazard Monitoring Group, Strada delle Cacce, 73, 10135 Torino, Italy

[15] Department of Agricultural, Forest and Food Sciences, University of Torino, Largo Paolo Braccini 2, 10095 Grugliasco (TO), Italy

[16] Master's degree in Sciences and technologies for environment and resources, University of Parma, Italy

[17] Agenzia Regionale per la Protezione dell'Ambiente della Valle d'Aosta ARPA, 11020 Saint-Christophe, Italy

[18] GLORIA Coordination, Institute for Interdisciplinary Mountain Research, Austrian Academy of Sciences & Center for Global Change and Sustainability, University of Natural Resources and Life Sciences Vienna (BOKU), Silbergasse 30/3, 1190 Vienna, Austria.

[19] Carabinieri, Biodiversity and Park Protection Dpt., Roma, Italy

[20] Department of Earth and Environmental Sciences, University of Pavia, via Ferrata 1, 27100 Pavia, Italy

[21] Lake and Glacier Research Group, Institute of Ecology, University of Innsbruck. Technikerstr. 25, 6020 Innsbruck, Austria

[22] Department of Environmental Sciences, University of Parma, Parco Area delle Scienze, 33/A, 43100 Parma, Italy

[23] Centre Alpien de Phytogéographie, Fondation J.-M. Aubert, 1938 Champex-Lac, Switzerland, & Section of Biology, University of Geneva, 1292 Chambésy, Switzerland

[24] Institute of Earth Surface Dynamics, University of Lausanne, Geopolis, 1015 Lausanne, Switzerland

[25] CNR ISAFOM, Ercolano (NA), Italy

**Abstract**

Mountain ecosystems are sensitive and reliable indicators of climate change. Long-term studies may be extremely useful in assessing the responses of high-elevation ecosystems to climate change and other anthropogenic drivers from a broad ecological perspective. Mountain research sites within the LTER (Long-Term Ecosystem Research) network are representative of various types of ecosystems and span a wide bioclimatic and elevational range.

Here, we present a synthesis and a review of the main results from long-term ecological studies in mountain ecosystems at 20 LTER sites in Italy, Switzerland and Austria covering in most cases more than two decades of observations. We analyzed a set of key climate parameters, such as temperature and snow cover duration, in relation to vascular species composition, plant traits, abundance patterns, pedoclimate, nutrient dynamics in soils and water, phenology and composition of freshwater biota.

The overall results highlight the rapid response of mountain ecosystems to climate change, with site-specific characteristics and rates. As temperatures increased, vegetation cover in alpine and subalpine summits increased as well. Years with limited snow cover duration caused an increase in soil temperature and microbial biomass during the growing season. Effects on freshwater ecosystems were also observed, in terms of increases in solutes, decreases in nitrates and changes in plankton phenology and benthos communities. This work highlights the importance of comparing and integrating long-term ecological data collected in different ecosystems, both terrestrial and freshwater, for a more comprehensive overview of the ecological effects of climate change. Nevertheless, there is a need for (i) adopting co-located monitoring site networks to improve our ability to obtain sound results from cross-site analysis, (ii)  carrying out further studies, in particular short-term analyses with fine spatial and temporal resolutions to improve our understanding of responses to extreme events, and (ii) increasing comparability and standardizing protocols across networks to clarify local patterns from global patterns.

1. **INTRODUCTION**

Mountains represent unique areas to detect climate change and assess climate-related impacts. One reason they are unique is that, as the climate rapidly changes with altitude over relatively short horizontal distances, so do vegetation and hydrology (Whiteman, 2000). Therefore, because of the complex topography in alpine environments, mountains exhibit high biodiversity (Winkler et al., 2016).

According to climate change projections, global warming will not be uniform but will vary considerably between different regions; in particular, climate change will be greater over land and at high latitudes and elevations (Auer et al., 2007; Gobiet et al., 2014). The high sensitivity of mountain areas with respect to climate change was clearly highlighted by the IPCC in its latest report (IPCC, 2014). Mountain ecosystems are indeed increasingly threatened by climate change, causing biodiversity loss, habitat degradation, deterioration of freshwater quality and landscape modifications (e.g., Körner, 2003), which poses a serious threat to the ecological integrity of terrestrial and freshwater ecosystems and the services they provide (Stoll et al., 2015; Huss et al., 2017). The response of mountain ecosystems may differ according to the rate of climate change, the ecological domain and the biogeographical region (Beniston, 2003; Müller et al., 2010).

High mountains in Europe contain 20% of the native flora of the continent (Väre et al., 2003) and are centers of plant diversity, hosting highly specialized vascular plants (Myers et al., 2000; Barthlott et al., 2005) and many endemic species (Langer & Sauerbier, 1997; Dirnböck et al., 2011; Stanisci et al., 2011). Climate change is considered one of the main threats to plant diversity above the tree-line. Recent model projections using climate change scenarios predicted a dramatic reduction of suitable habitats for high-elevation herbaceous plants (Engler et al., 2011) even if thermal microhabitat mosaics offer alpine species both refuge habitats and serve as stepping stones as atmospheric temperatures rise (Scherrer, Körner 2011). The slow growth rates of long-lived alpine plants may lead to a delayed decrease in the ranges of species, creating an extinction debt (Dullinger et al., 2012). Mountain forests are also particularly vulnerable to climate change due to their long rotation cycles that may hinder their adaptation capacity (Lindner et al., 2010). However, forests can also benefit from global change, as increasing concentrations of $CO_2$ and nitrogen deposition should increase photosynthesis rates and forest growth (e.g., Matyssek et al., 2006).

Freshwater ecosystems in mountain areas are of paramount importance as high-quality water resources and biodiversity hotspots because they host specialized aquatic biotas (Körner, 2004). Mountain lakes are particularly sensitive to the effects of global change, such as the deposition of atmospheric pollutants and increasing temperatures (Battarbee et al., 2009; Catalan et al., 2009). Mountain lakes are usually small (< 0.5 km$^2$), relatively shallow and generally ice-covered for prolonged periods (from 3-4 to 8-9 months per year). Organisms living in these lakes face harsh environmental conditions, low nutrient availability, and extreme changes in light conditions during the year (Sommaruga, 2001). Riverine habitats in glacier catchments are also among the most vulnerable habitats with respect to climate change. River community structure (species composition, abundance, and ecological traits) is related to geomorphological features, which are in turn affected by glacier dynamics (Füreder, 2007; Finn et al., 2010) that are strongly affected by climate change.

Increased air temperatures due to climate change resulted in shorter snow cover seasons due to later accumulation and earlier meltdown (Klein et al., 2016). A general decrease in the spatial extent of spring snow cover in the Northern Hemisphere has been reported (IPCC, 2014) as well as an upward shift of the rain–snow line (Lundquist et al., 2008). In the Alps, mean snow depth, snow cover duration and number of snowfall days have decreased since the early 1980s, although with large regional and altitudinal variations (Laternser and Schneebeli, 2003). Changes in snow cover may in turn affect mountain ecosystem hydrology (Gobiet et al., 2014), biogeochemical processes in soil and water (Magnani et al., 2017), plant composition, phenology and structure (Grabherr et al.

1995). The insulating properties of snow influence the underlying soil temperature regime and the extent to which soil is directly exposed to cold air temperatures in the winter (Edwards et al., 2007). Indeed, mountain soils typically experience freezing conditions only during the early winter. After the deposition of snow cover that insulates the underlying soils from low air temperatures, soils remain unfrozen during most of the winter season (Jones, 1999).

A valuable volume of long-term ecological data that can be used for defining and testing the consequences of climate change on mountain ecosystems is available from the LTER Network (Long Term Ecosystem Research; http://www.lter-europe.net/). The LTER is an international monitoring network that gathers multiyear high-quality ecological data that are periodically collected to assess the impacts of global change on ecological processes. The operation of several LTER sites distributed along the Apennines and the Alps offers an excellent resource to develop and test the effects of climate change on different types of mountain ecosystems. Indeed, long-term ecosystem research has been successfully carried out adressing the impact of climate change in the different high elevation ecosystems, and very interesting results have been found (Müller et al., 2010). However, these results have often been determined only at a site-specific or regional level. Based on the urgent demands for long-term research with comparative ecological analysis (Müller et al. 2010), a comprehensive overview accounting for the effects of climate change in different mountain ecosystems needs further attention and research efforts.

In this context, this paper presents a new analysis of existing ecological data for many LTER sites, both terrestrial and freshwaters, and aims to summarize the complex information obtained from the long-term observations of different mountain ecosystems in response to climate change. We aimed to assess and possibly synthetize the response of the main ecological processes in different mountain ecosystems to climate change. In particular, the ecological changes we tested with our analyses were: (i) the change in vegetation cover and C-uptake; (ii) the alteration of biogeochemical cycles in soils and water; and (iii) the change in phenology and biological diversity.

We focused on the following compartments and ecological parameters:

- soil: the interannual variability in soil temperature and nutrient cycling was investigated at one LTER site in the Alps in relation to snow cover duration and pedoclimatic conditions;
- vegetation: we assessed changes in vegetation cover over time considering both regional scale and elevation belts, using newly collected data from LTER high-elevation sites;
- freshwater: we assessed interannual variability and long-term changes in selected chemical and biological variables in response to climate drivers at some LTER lake and river sites representative of different mountain areas.

In addition, we discuss previous and on-going studies in LTER mountain sites dealing with the effects of climate drivers on additional ecological parameters, including grassland ecosystem productivity, forest carbon storage and animal population dynamics.

## 2. RESEARCH SITES AND DATA ANALYSIS

### 2.1 Study area and climate driver description

We considered 20 research sites, representative of the Alps, from west to east (Italy, Switzerland and Austria) and of the Apennines (Italy) (Tab. 1, Fig. 1).

The sites are located between 1300 and 3212 m a.s.l. The sites are not directly affected by anthropogenic disturbance, or are natural sites under sustainable management or under mild pressure (e.g., land abandonment, reduced grazing, sustainable forest management). Fourteen sites are terrestrial sites (forests, grasslands, alpine tundra and nival areas), and six are freshwater sites (lakes and rivers). Details on site characteristics, the purpose and history of the research site and on-going studies can be found at the specific links in the DEIMS-DSR portal (Dynamic Ecological Information Management System - Site and dataset registry; https://data.lter-europe.net/deims/) provided in Tab. 1. Long-term ecological data collected at the selected sites were used in this paper for assessing decadal changes in soil properties, vegetation cover and chemical and biological features of freshwater ecosystems.

Tab. 1 - LTER mountain sites considered. June anomaly refers to the air temperature anomaly (1995-2015) with respect to a 1961-1990 base period (see Fig. 1).

| Site code | Site name | Elevation - average m a.s.l. | June anomaly | Site type | Parent site | Established | Site description in DEIMS |
|---|---|---|---|---|---|---|---|
| LTER_EU_IT_021 | Central Apennines: Gran Sasso d'Italia | 2210 | 1.46 | Terrestrial | IT01- Apennines - High elevation Ecosystems | 1986 | https://data.lter-europe.net/deims/site/lter_eu_it_021 |
| LTER_EU_IT_025 | Central Apennines: Velino Duchessa (VEL) | 2145 | 1.36 | Terrestrial | IT01- Apennines - High elevation Ecosystems | 1993 | https://data.lter-europe.net/deims/site/lter_eu_it_025 |
| LTER_EU_IT_022 | Central and southern Apennines: Majella-Matese (MAJ -MAT) | 2400 | 1.22 | Terrestrial | IT01- Apennines - High elevation Ecosystems | 2001 | https://data.lter-europe.net/deims/site/lter_eu_it_022 |
| LTER_EU_IT_023 | Northern Apennines (NAP) | 1900 | 3.02 | Terrestrial | IT01- Apennines - High elevation Ecosystems | 2001 | https://data.lter-europe.net/deims/site/lter_eu_it_023 |
| CH-VAL | W-Alpes: Alps of Valais-Entremont (Switzerland) (VAL) | 2777 | 1.11 | Terrestrial | Project GLORIA site, not yet in LTER network | 2001 | |
| IT-ADO | S-Alps, Dolomites (Italy) (ADO) | 2705 | 1.45 | Terrestrial | Project GLORIA site, not yet in LTER network | 2001 | |
| LTER_EU_IT_073 | W-Alpes: Mont Avic (MAV) | 2340 | 1.41 | Terrestrial | IT19 - High elevation sites in the NWAlps | 2001 | https://data.lter-europe.net/deims/site/lter_eu_it_073 |
| LTER_EU_AT_007 | E-Alps: Hochschwab (HSW) | 2100 | 0.85 | Terrestrial | LTES Platform Eisenwurzen (EW) | 1998 | https://data.lter-europe.net/deims/site/lter_eu_at_007 |
| LTER_EU_IT_074 | W-Alps: Cime Bianche | 3100 | 1.30 | Terrestrial | IT19 - High elevation sites in the NW Alps | 2006 | https://data.lter-europe.net/deims/site/lter_eu_it_074 |
| LTER_EU_IT_077 | W-Alpes: Torgnon grassland Tellinod | 2100 | 1.30 | Terrestrial | IT19 - High elevation sites in the NW Alps | 2008 | https://data.lter-europe.net/deims/site/lter_eu_it_077 |
| LTER_EU_IT_076 | W-Alpes: Istituto Scientifico Angelo Mosso | 2700 | 1.28 | Terrestrial | IT19 - High elevation sites in the NW Alps | 1928 | https://data.lter-europe.net/deims/site/lter_eu_it_076 |
| LTER_EU_IT_031 | Central Apennines: Collelongo-Selva Piana ABR1 | 1500 | 1.36 | Terrestrial | IT03-Forest of the Apennines | 1991 | https://data.lter-europe.net/deims/site/lter_eu_it_031 |
| LTER_EU_IT_033 | Central Apennines: Montagna di Torricchio | 1260 | 0.99 | Terrestrial | IT03-Forest of the Apennines | 1971 | https://data.lter-europe.net/deims/site/lter_eu_it_033 |
| LTER_EU_IT_109 | W- Alps: Gran Paradiso National Park | 2500 | 1.34 | Terrestrial | IT23 - Gran Paradiso National Park - Italy | 1922 | https://data.lter-europe.net/deims/site/lter_eu_it_109 |
| LTER_EU_IT_089 | W-Alps: Lake Paione Superiore | 2269 | 1.41 | Freshwater | IT09-Mountain Lakes | 1978 | https://data.lter-europe.net/deims/site/lter_eu_it_089 |
| LTER_EU_IT_088 | W-Alps:Lake Paione Inferiore | 2002 | 1.41 | Freshwater | IT09-Mountain Lakes | 1978 | https://data.lter-europe.net/deims/site/lter_eu_it_088 |
| LTER_EU_IT_047 | Northern Apennines:Lake Scuro Parmense | 1527 | 2.21 | Freshwater | IT09-Mountain Lakes | 1986 | https://data.lter-europe.net/deims/site/lter_eu_it_047 |
| LTER_EU_IT_046 | Northern ApenninesLake Santo Parmense | 1507 | 2.21 | Freshwater | IT09-Mountain Lakes | 1952 | https://data.lter-europe.net/deims/site/lter_eu_it_046 |
| LTER_EU_AT_012 | E-Alps: Gossenköllesee | 2417 | 1.24 | Freshwater | LTSER Platform Tyrolean Alps (TA) | 1975 | https://data.lter-europe.net/deims/site/lter_eu_at_012 |
| LTER_EU_IT_100 | E-Alps: Saldur River | 2000 | 1.46 | Freshwater | IT25 - Val Mazia/Matschertal | 2008 | https://data.lter-europe.net/deims/site/lter_eu_it_100 |

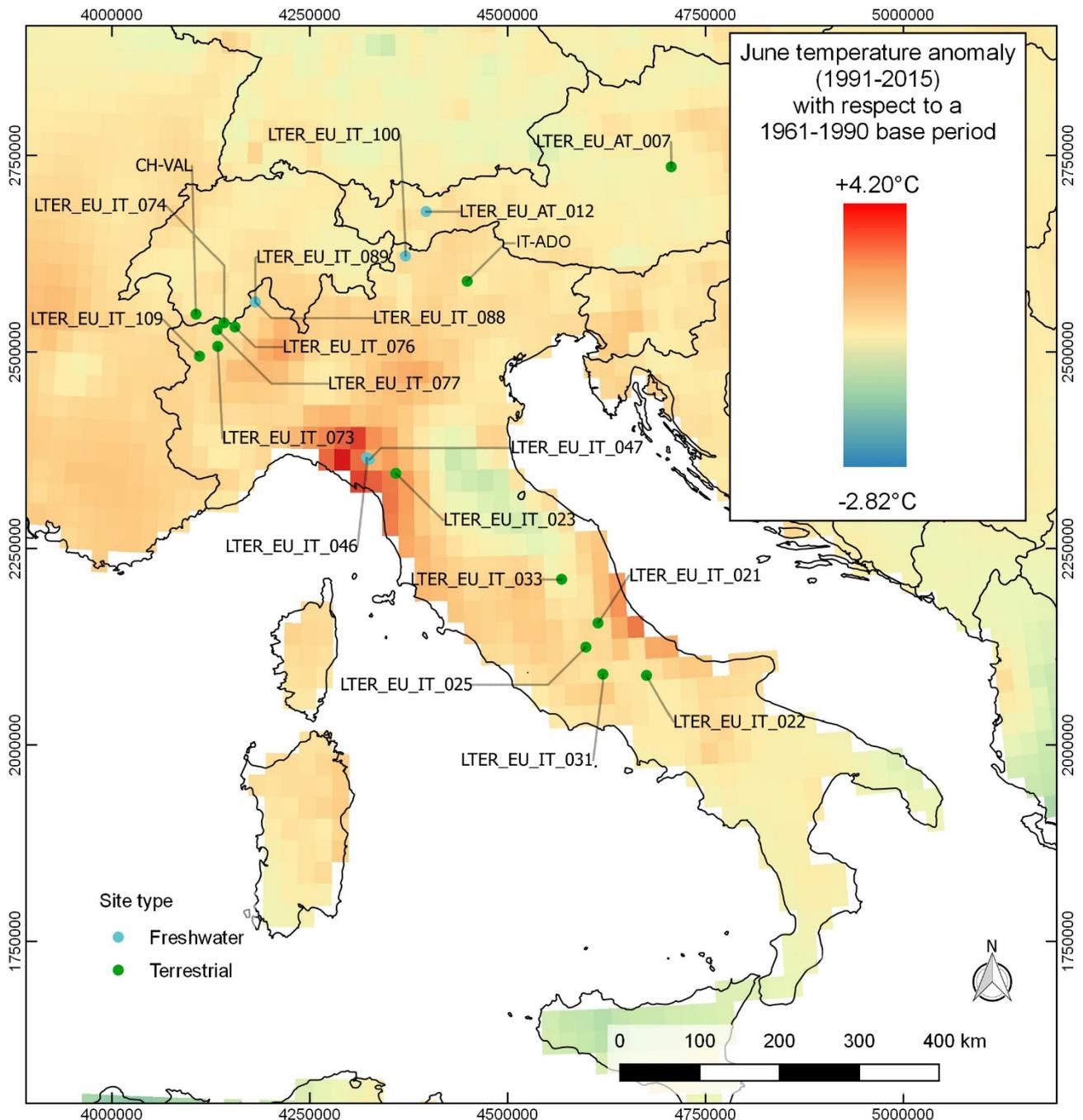

Fig. 1. Location of the LTER sites. The map shows the June temperature anomaly (1995-2015) with respect to a 1961-1990 base period (map prepared from data provided by E-OBS, Haylock et al. 2008, resolution 0.25°; Coordinates are in ETRS-LAEA/ETRS89 – EPSG: 3035).

In the Northern Hemisphere, the IPCC (2014) indicated that the period between 1983 and 2012 was the warmest of the millennium. Moreover, the IPCC reported recent changes in precipitation patterns due to an increase in the atmospheric moisture content.

Extensive studies on alpine climate showed that air temperature increases were quite homogeneous over the Alps (EEA, 2009). The long-term climatic data available for the alpine tundra (LTER site

LTER_EU_IT_076 "Istituto Mosso", time span 1928-2013, station "Gabiet") showed an increase in maximum air temperature equal to 0.015°C y$^{-1}$ (Fratianni et al., 2015). In the Ossola Valley, Western Alps, where some LTER lake sites are located (LTER_EU_IT_088, LTER_EU_IT_089), long-term air temperature data (since the 1930s) showed an average increase rate of 0.011°C y$^{-1}$ (0.015°C y$^{-1}$ in the summer) (Rogora et al., 2004). In the central Apennines, an average temperature increase rate of 0.027°C y$^{-1}$ occurred during the period 1950-2014 (Evangelista et al., 2016).

The research sites considered in this paper all experienced an increase in air temperature over the past two decades, as shown by the June temperature anomaly (1991-2015) with respect to a 1961-1990 base period (Fig. 1). The increments range from 0.85°C at Hochschwab (LTER_AT_007) to 3.02°C in the Apennines (LTER_IT_023). We considered June temperatures because June represents a key month for mountain ecosystem phenology. For instance, June coincides with the first part of the growing period, which is the most relevant for plant growth (Gottfried et al., 2012). Moreover, at high elevations in the Alps and Apennines, June temperatures strongly influence snow melting rates, with significant effects on the beginning of the growing season (e.g., Sedlacek et al., 2015). June temperatures may also be critical for high-elevation lakes and rivers, because they also affect the timing of ice-break (Preston et al., 2016) and can drive the onset of lake water stratification and stream water discharge. Finally, June represents a crucial month for many mountain vertebrates, including ungulates, because it corresponds to the birthing season and, at lower elevation (1300 – 1800 m a.s.l.), this month is also crucial for forest carbon sequestration.

Precipitation trends are more spatially variable compared to trends in air temperature (IPCC, 2014). Climate warming is predicted to cause changes in the seasonality of precipitation, with an expected increase in intra-annual variability, more intense precipitation extremes, and more potential for flooding (Gobiet et al., 2014). Projected changes in precipitation, snow cover patterns and glacier storage in the Alps will also alter runoff regimes, leading to more droughts in the summer (EEA, 2009).

Snow cover duration (SCD) is also an important driver of change in mountain ecosystems. A reduction in the snow cover amount and extent has been described for areas in the Alps below 2000 m a.s.l. Specifically, the delayed onset of snow and the anticipated snowmelt contribute to an overall decrease in SCD (Klein et al., 2016). A decrease in snow cover depth and SCD has been specifically recorded in some of the analyzed LTER lake sites (Western Alps) over the last 30 years (Rogora et al., 2013).

## 2.2 Data collection and analysis

The data analyses were partly performed on already existing datasets, developed in the framework of previous and on-going research projects (e.g., GLORIA: GLobal Observation Research Initiative in Alpine environments; http://www.gloria.ac.at), and partly on datasets specifically developed for this paper (e.g., lake chemistry). Information about the datasets for each site, their availability and the methods used for generating the data are provided in DEIMS-DSR (Tab. 1). Some of the sites included in our analysis are also part of the NEXTDATA special project "Data-LTER-Mountain" (http://www.nextdataproject.it/). Within this project, a distributed system of archives and access services to data and metadata of the Italian LTER sites located in mountain regions has been developed.

### 2.2.1 Snow cover duration and soil properties

To analyze the regional pattern of SCD and of the snow melting date we used soil temperatures collected at different LTER research sites (LTER_EU_IT_031, LTER_EU_IT_022, LTER_EU_IT_077 and LTER_EU_IT_076). We used thermistors combined with data loggers placed at a soil depth of 10 cm for the measurement of the hourly soil temperature (instrument sensitivity of ± 0.1°C). The SCD at each study site was calculated based on the daily soil temperature amplitude (Schmid et al. 2012). When the daily soil temperature amplitude remained within a range of 1 K, the day was defined as a "snow cover day" (Danby and Hik, 2007). The SCD was calculated as the sum of the snow-covered days. When the daily mean soil temperature dropped below and rose above 0°C, it was considered as a freeze/thaw cycle (FTC) (Phillips and Newlands, 2011) that approximately corresponded to the melting period. Then, the snow melting date (DSO1: day since October 1$^{st}$) was assessed.

To assess the impact of SCD changes on soil properties, i.e., temperature and moisture during the growing season, microbial carbon, microbial nitrogen, ammonium, nitrate, dissolved organic carbon and dissolved organic nitrogen, we used both meteorological and physico-chemical soil data recorded in 2008-2016 at the LTER site LTER_EU_IT_076, located close to Monte Rosa Massif. The study was conducted at three high-elevation subsites located in the upper part of a glacial valley, at an elevation ranging between 2500 and 2800 m a.s.l. (Table S1). Each subsite consisted of three plots. Soil temperatures collected by data loggers and thermistors were used for calculating the beginning of the growing season that was defined as the time when weekly topsoil temperature reached 3°C (Paulsen and Körner, 2014). For soil characteristics, a soil sampling campaign was performed in mid-September, approximately at the end of the growing season. Every year (2008-2016), three soil samples (A horizon, 0–10 cm depth) were collected at each subsite, which in turn consisted of three subsamples in each plot. Samples were homogenized by sieving at 2 mm within 24 h of collection. At each sampling time, subsamples were dried at 100°C overnight to obtain the gravimetric water content. An aliquot of 20-g of fresh soil was extracted with 100 mL $K_2SO_4$ 0.5 M as described by Brooks et al. (1996), whereas a 10-g aliquot was subjected to chloroform fumigation for 18 h before extraction with 50 mL of $K_2SO_4$ 0.5 M. Dissolved organic carbon was determined with 0.45 μm membrane-filtered $K_2SO_4$ extracts (extractable DOC) with a total organic carbon (TOC) analyzer (Elementar, Vario TOC, Hanau, Germany). Microbial carbon (Cmicr) was calculated from the difference in DOC between fumigated and non-fumigated samples that were corrected by a recovery factor of 0.45 (Brookes et al., 1985). The ammonium in the $K_2SO_4$ extracts (extractable $N-NH_4^+$) was diffused into a $H_2SO_4$ 0.01M trap after treatment with MgO (Bremner, 1965), and the trapped $NH_4^+$ was determined colorimetrically (Crooke and Simpson, 1971). The nitrate (extractable $N-NO_3^-$) concentration in the same extracts was determined colorimetrically as $NH_4^+$ after reduction with Devarda's alloy (Williams et al., 1995). Total dissolved nitrogen (TDN) in the extracts was determined as reported for DOC. Dissolved organic nitrogen (extractable DON) was determined as the difference between TDN and inorganic nitrogen ($N-NH_4^+ + N-NO_3^-$) in the extracts. Microbial nitrogen (Nmicr) was calculated from the difference in TDN between the fumigated and non-fumigated samples that were corrected by a recovery factor of 0.54 (Brookes et al., 1985).

### *2.2.2 Vegetation cover*

We used vegetation data that were collected in 2001 and 2015 in permanent plots at eight high mountain study areas of the LTER and GLORIA networks and distributed in the Italian Apennines and Alps and in the Austrian and Swiss Alps (Tab. S2). According to the GLORIA sampling design (Pauli et al., 2015), each site comprises 2-4 summits along an elevation gradient. For each summit, a 3×3 $m^2$ grid was established for each cardinal direction at 5 m below the summit peak. In the four 1-$m^2$ corner plots (4 quadrats) of the grids, the percentage cover for each plant species was estimated. At one site (Velino: VEL), the plant species cover was collected on permanent plots placed along an elevational gradient that was divided into six 100 m elevation bands (from 1800 to ~ 2400 m a.s.l.; Theurillat et al., 2007). For each band, four to six 2 m x 2 m vegetation plots were sampled.

We analyzed vegetation cover change over time (T1 – T2) for mountain sites (i.e., considering all the plots grouped by site) and vegetation belts (i.e., grouping plots by vegetation belt). The vegetation belts present in the analyzed sites were defined according to Pignatti (1979) and Theurillat et al. (1998).

We analyzed changes in vegetation cover by first calculating the total vegetation cover per plot as the sum of the cover estimates of the individual plant species as a proxy, non-destructive measure, of aboveground biomass (Fry et al., 2013). Then, we quantified vegetation change for each site and vegetation belt, by calculating the effect size obtained by computing the weighted average of the standardized difference (based on pooled variance measures) between the mean cover values on the two sampled dates (T1 and T2; unbiased estimator Hedge's g; Hedges, 1981). This standardized difference, which estimates the effect size as the difference between T2 and T1, provides an estimate of the magnitude of an effect, i.e., the cover change between the sampled dates, when data collection varies among studies (e.g. Elmendorf et al., 2012). The effect size is positive when the vegetation cover increases over time and is negative when the cover decreases. We randomly chose one of the four plots in each cardinal direction for the GLORIA sites and one from each band for the Velino site. For each randomly extracted plot, we calculated the difference between cover values on the two dates (T2-T1), and based on these differences, we computed the median and the 95% confidence intervals of Hedge's g by using the BootES package (Kirby and Gerlanc, 2013) in the R statistical software (R Development Core Team, 2011). In addition to the weighted median effect size, we also reported the median percent change in all the studies.

### 2.2.3 *Freshwater chemistry and biology*

The high elevation LTER lakes considered in this paper (Tab. S3) are located both in the Alps (Italy and Austria) and in the Apennines (Central Italy). They have different origin, morphometry and surrounding land cover characteristics. However, they share some common features, such as an oligotrophic status, diluted waters with low solute content and relatively simple food webs compared to lowland lakes (Rogora et al., 2013; Sommaruga 2001). To test for common trends duting 1980-2016, the following variables were considered: conductivity, alkalinity, sulphate and nitrate ions. Because data have been collected with different sampling methods (e.g., vertical profiles at the deepest points, sampling of lake outlets or at the lake shores) and frequency, we selected surface data representative of the late summer or early autumn period, when lakes are more stable and uniform from a chemical point of view. Chemical analyses were performed according to standard methods for freshwater (APHA AWWA WEF, 2012). We assessed trend significance and slope by the Mann-Kendall test (Hirsch et al., 1982) and Sen's method (Sen, 1968), respectively.

In addition to chemical trends, long-term biological data available for two lakes in the Apennines were analyzed (SCU and SAN; Tab. S3). The two lakes are characterized by different size, trophic structure, and level of anthropogenic disturbance. Extensive information on the lakes' chemico-physical and biological characteristics can be found in Bondavalli et al. (2006) and Bertani et al. (2016). Monthly values of water temperature throughout the water column, chlorophyll-a concentrations and zooplankton species abundance during the open-water season (May-October) are available for these two lakes for the following periods: SAN, 1971-1975 and 2012 and SCU, 1986-2012. Monthly average air temperatures duting 1971-2012 were derived from a weather station near the two lakes (Passo della Cisa: 44°26' N, 9°25' E) (data downloaded from the National Oceanic and Atmospheric Administration (NOAA) website; www.climate.gov). From the same website, we also downloaded monthly values of the East Atlantic pattern (EA) climatic index, a teleconnection that influences climate in the Mediterranean region throughout the year (Kutiel and Benaroch, 2002). For both air temperature and EA we calculated seasonal averages for the winter (December through February), spring (March through May), summer (June through August) and autumn (September through November). We hypothesized that positive values of the EA climatic index would be associated with higher air and water temperatures and earlier seasonal plankton development.

For the two lakes, we analyzed interannual changes in the phenology of chlorophyll-a (a proxy for phytoplankton abundance) and zooplankton taxa by calculating different phenology metrics. Specifically, we estimated the timing of the seasonal phytoplankton peak each year by calculating the "center of gravity" for monthly chlorophyll-a values across the open-water season (Edwards and Richardson, 2004; Thackeray et al., 2012). For each of the dominant zooplankton taxa, we characterized population phenology each year by calculating the date of its first appearance and the date of its peak population abundance (Adrian et al., 2006).

We tested for relationships between large-scale climatic patterns (EA) and interannual variability in both local climate features (air and water temperature) and plankton phenology by calculating correlations (Spearman's correlation coefficient) between the average seasonal EA values and 1) corresponding average seasonal air and water temperature values and 2) phyto- and zooplankton phenology metrics.

Benthic community structure was evaluated at the LTER river site LTER_EU_IT_097 Matsch/Mazia Valley (Tab. S3) in the upper Vinschgau/Venosta Valley (South Tyrol, Italy). The Saldura Stream, draining the Matsch/Mazia Valley, one of the driest valleys of the Alps, represents an ideal site to focus on climate change impacts. The Matsch/Mazia Valley is characterized by the presence of a glacier that is rapidly melting. The glacier extends from 2,800 m to 3,500 m a.s.l. To evaluate the influence of the glacier and the spatial patterns of the macrobenthic assemblage, three sampling stations were selected along the main stream at increasing distances from the source (located from 2400 m a.s.l. to 1500 m a.s.l.). The macrobenthic community was analyzed by applying the multi-habitat sampling methodology using a standard Surber Sampler. The biological samples were integrated by chemico-physical analyses of the running water to correlate community composition and diversity with environmental variables. We assessed temporal and spatial distribution patterns of the macroinvertebrate community and related the biological results with the changing abiotic conditions.

## 3. RESULTS AND DISCUSSION

### 3.1 Snow cover duration and effects on soil properties and animal population dynamics

The SCD and melting date calculated from the soil temperature data greatly varied across elevational and latitudinal gradients, from short SCD (~100 days) and early snow melting dates (DSO1~ 200) in the lower altitudes (~1500 m a.s.l.) and latitudes (Lat. 41 N) to long SDC (~250 days) and late snow melting dates (DSO1~300) in the higher altitudes (~2800 m a.s.l.) and latitudes (Lat. 46 N) (Fig. 2). Our results agree with previous studies reporting a strong relation between the SCD and the snow melt day for the Swiss Alps (Klein et al., 2016).

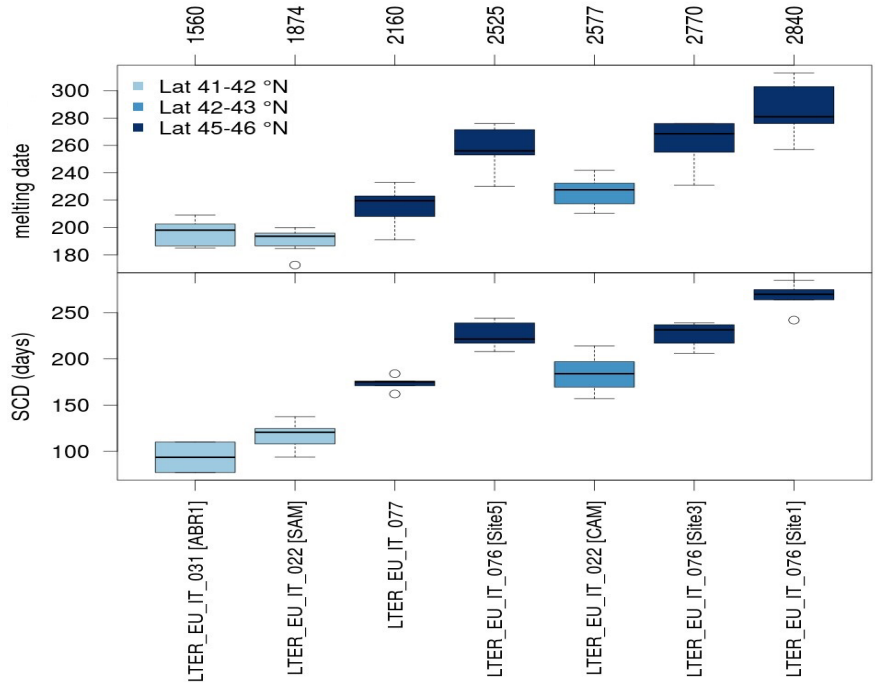

Fig. 2 - Melting dates in DSO1 (Day Since October 1st) (above) and SCD (days) (below) calculated from soil temperature data at selected LTER sites across elevation and latitude gradients. Yearly average values for the period 2008-2015. For more details about analyzed LTER sites see Tab. 1.

The analysis of soil temperature data collected at the research site Istituto Scientifico Mosso (LTER_EU_IT_076) demonstrated that the mean soil temperature of the snow cover season is -0.19°C (±0.31). The beginning of the snow cover season showed lower temporal variability than the end of the SCD, ranging between October 27 (±15 days) and June 24 (±20 days), respectively.

The analysis of the relation among SCD and both soil temperature and microbial biomass recorded during the growing season showed the occurrence of significant negative correlations (r=-0.621, p<0.01 and r=-0.566, p<0.01, respectively; Fig. 3). As observed by Magnani et al. (2017), a short SCD may increase soil temperature and substrate availability during the subsequent growing season, favoring soil microbial biomass.

Soil $N-NH_4$ (r=-0282; p<0.05), DOC (r=-0.427; p<0.01) and $N_{micr}$ (r=-0.403; p<0.01) were inversely correlated with SCD, while $N-NO_3$ showed a sharp increase during 2008-2009, characterized by rather extreme meteorological conditions; in particular, the 2007-2008 winter season was characterized by a thinner snow depth (max approximately 200 cm) than the average snow depth value (289 cm, time-span 2008-2016), while the 2008-2009 winter season had a thicker snow depth (max of 560 cm). The little snowpack in 2007-2008 caused a large number of soil FTCs, which could have contributed to the destruction of the soil aggregates and the release of previously unavailable organic N (Freppaz et al., 2007). The thick snowpack recorded during the next winter season could have released a greater N-NO$_3$ input into the soil than average during snow melt.

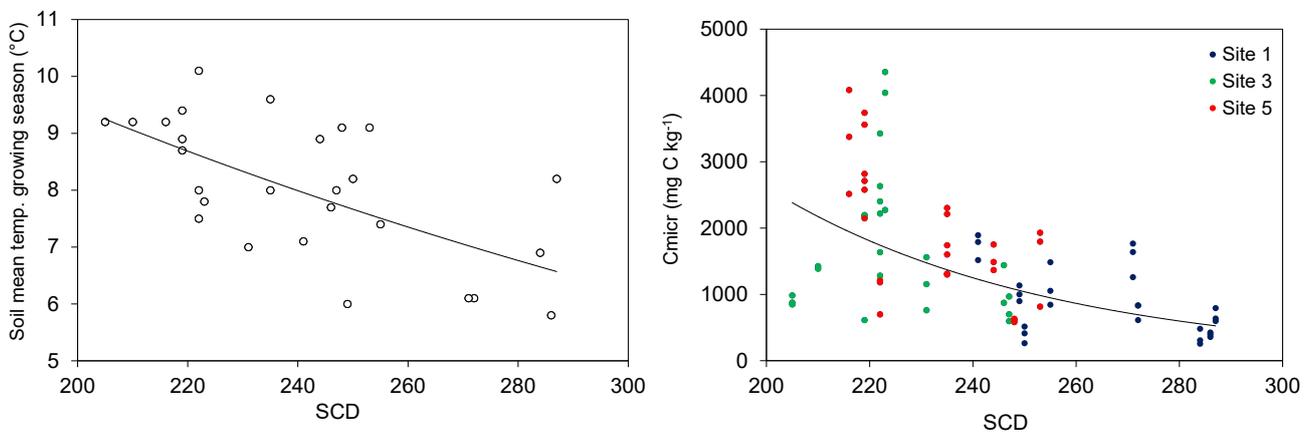

Fig. 3 - Scatterplots between SCD and soil mean temperature ($r = –0.62$, $p < 0.01$) (left panel) and between SCD and C$_{micr}$ (r=-0.57, $p < 0.01$) (right panel) measured in the growing season considering all the study subsites of LTER_EU_IT_076 (site 1, site 3, and site 5; Table S1) during 2008-2016 ($n = 79$).

Snow cover is also an important driver for animal population dynamics. Long-term studies that have been performed at the LTER site LTER_EU_IT_109 Gran Paradiso National Park since the 1950s showed how the dynamics of the Alpine ibex (*Capra ibex*) prior to 1980s was mainly driven by the average winter snow depth that represented a limiting factor for population growth (Jacobson et al., 2004). In general, the snow cover effect on Alpine ibex survival was not linear and not equal for all sex and age classes, and the snow effect was amplified during years of high animal densities (Mignatti et al., 2012). Snow precipitation patterns proved to also be important for vegetation growth and consequently for resource availability (Pettorelli et al., 2007).

## 3.2 Vegetation cover at high elevations and carbon sequestration in mountain forests

The analysis of vegetation cover changes showed positive effect size estimates for all high mountain sites (Fig. 4) but with site-specific magnitudes. In particular, the Majella site (CAM) had the largest positive and significant effect size (Hedge's g=0.70, 95% CI, 1.45 – 0.08), followed by Dolomites (ADO) (Hedge's g=0.55, 95% CI, 0.86 – 0.29).

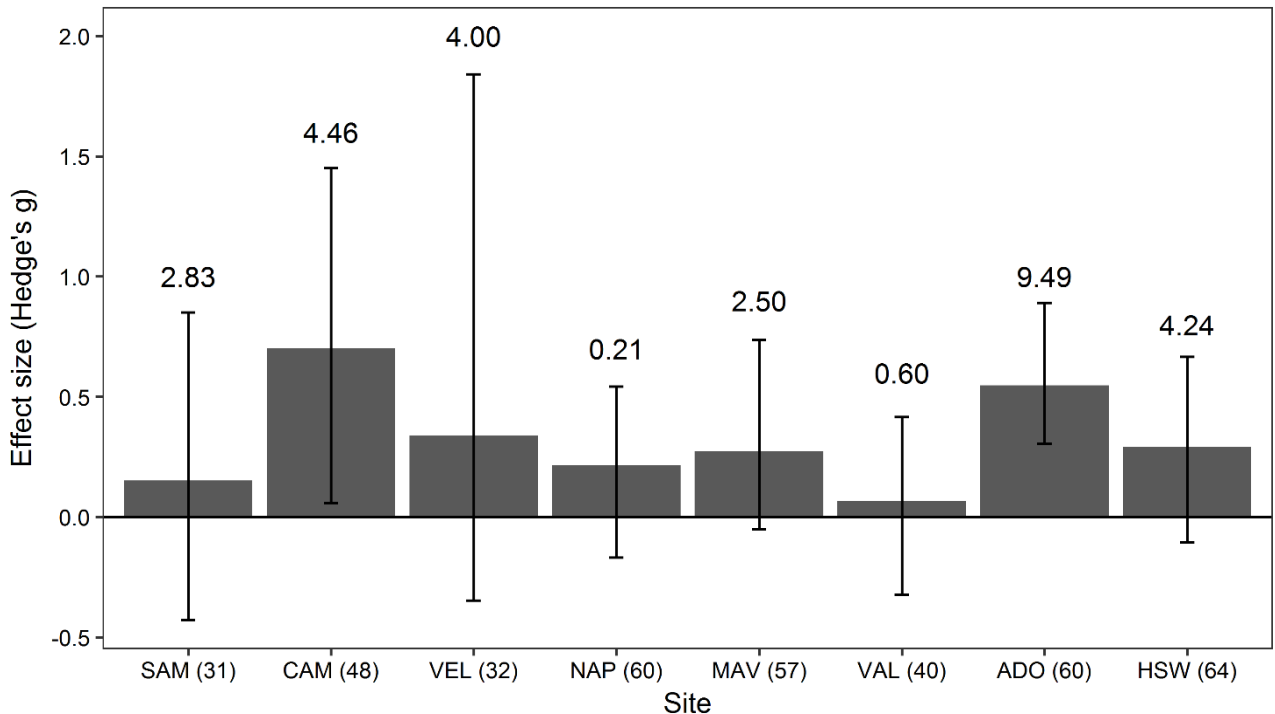

Fig. 4. Median effect sizes (Hedge's g) of temporal change on vegetation cover for each site separately (sites are arranged from south to north). Error bars show 95% confidence intervals. An effect size is significantly different from zero when its 95% confidence interval does not overlap zero. Median percent change recorded over all plots is shown above the corresponding bar. Abbreviations refer to Table 1. The number of plots per site is indicated in parenthesis.

A general tendency towards increased vegetation cover was also observed in Matese (SAM), Velino (VEL), Northern Apennines (NAP), Mont Avic (MAV), Hochschwab (HSW) and Valais-Entremont (VAL).

The analyses of plant cover changes per vegetation belt showed a positive effect size for treeline, subalpine and alpine belts and no effect for the nival belt (Fig. 5). The increase in plant cover over the last fifteen years is most likely related to a greening process (Carlson et al., 2017), which reduced vegetation gaps and was promoted by the expansion of the most thermophilic species already present in the plots and/or to the immigration of species from lower elevations (Gottfried et

al., 2012).

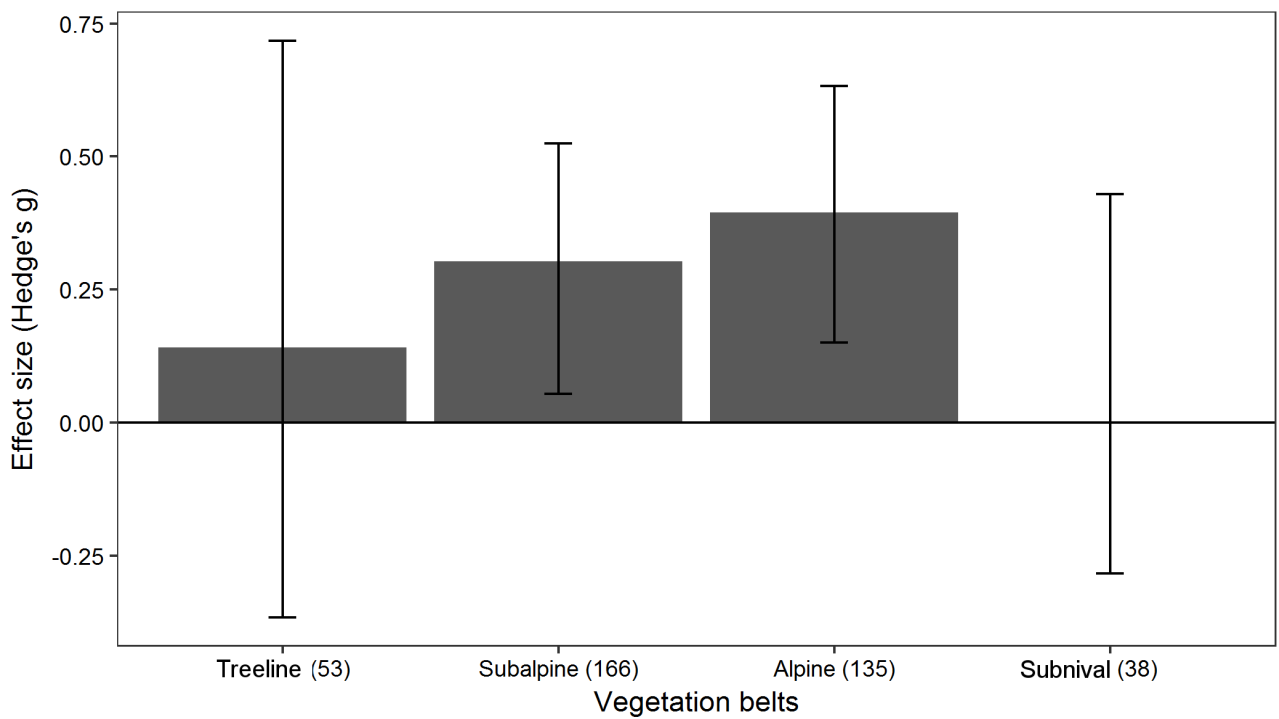

Fig. 5. Median effect sizes (Hedge's g) of temporal change on vegetation cover grouped by vegetation belt. Error bars show 95% confidence intervals. An effect size is significantly different from zero when its 95% CI does not overlap zero. The number of plots per vegetation belt is indicated in parenthesis.

Indeed, global warming affects high mountain ecosystems by increases in temperature, early snowmelt and a prolonged growing season (Pauli et al., 2012). These factors might have played a key role in the observed increase in plant cover. In fact, the air temperatures before snowmelt and after the meltdown (i.e., the May/June temperatures) are the main factor affecting plant growth in these ecosystems (Jonas et al., 2008; Rammig et al., 2010; Carbognani et al., 2016). However, climate change probably interacted with land-use change, which can exacerbate the effects of climate warming on mountainous vegetation (Theurillat and Guisan, 2001), although such an issue is still largely unexplored (Chelli et al., 2017).

On European mountain summits, increasing atmospheric temperatures already have resulted in a measurable expansion of thermophilic species that increased their cover in situ and migrated from lower elevations into the alpine lifezone (Grabherr et al., 1995; Gottfried et al., 2012; Jiménez-Alfaro et al. 2014). Our results agree with recent local research that provided evidence of an increase in caespitose hemicryptophytes and suffruticose chamaephytes frequencies on CAM summits (Stanisci et al., 2016) and an overall increase in species frequency on ADO summits (Erschbamer et al., 2011; Unterluggauer et al., 2016).

The temporal analysis of vegetation belts at the LTER and GLORIA mountain sites highlighted a significant increment in vegetation cover at these alpine sites, followed by the subalpine ones, whereas sites located at the treeline belt showed negligible variation, and vegetation cover at the nival belt did not show any change. The specific behavior of vegetation cover on each elevational belt probably reflects the natural structure patterns of plant communities across the elevation gradient. For instance, the cover of subalpine and alpine sparse/open vegetation in ridge habitats significantly increased, appearing more prone to a greening process. This process is likely due to the expansion of plant species already present at the site and to colonization events, which have been recorded in previous papers based on long-term vegetation analysis in alpine environments (Walther et al., 2005; Vittoz et al., 2009b; Matteodo et al., 2016; Carbognani et al., 2014).

Conversely, although nival ridges experienced changes in species composition (e.g., Pauli et al. 2012), this did not seem to result in an increase in vegetation cover. The environmental constraints at these elevations likely do not allow a greening process, at least until now, but only plant species turnover.

Previous vegetation studies carried out at high-elevation LTER and GLORIA sites in the Alps and Apennines indeed revealed changes in plant community structure and composition. Specifically, an increase in species richness (Erschbamer et al., 2008; Erschbamer et al., 2011; Pauli et al. 2012; Unterluggauer et al., 2016) and changes in community compositions (Petriccione, 2005; Erschbamer et al., 2008; Erschbamer et al., 2011; Stanisci et al. 2016) were observed in response to temperature increases and changes in precipitation patterns (Petriccione, 2005). For instance, Unterluggauer et al. (2016) observed a 9% to 64% increase in species richness over a period of 14 years in the Southern Alps (Dolomites). Similarly, Petriccione (2005) observed a 10 to 20% increase in species richness at the research site Gran Sasso d'Italia (LTER_EU_IT_021) over a period of nine years.

As observed on other European summits (Gottfried et al., 2012), a thermophilization process also occurred at the Apennine and Alps LTER and GLORIA sites (Theurillat and Guisan, 2001; Pauli et al., 2007; Holzinger et al., 2008; Parolo and Rossi 2008; Engler et al., 2011, Erschbamer et al., 2008, 2011; Matteodo et al., 2013; Cannone and Pignatti, 2014; Stanisci et al., 2016) and the arrival of new species typical of lower elevation belts was registered (Petriccione, 2005; Vittoz et al., 2008; Erschbamer et al., 2009; Pauli et al., 2012; Evangelista et al., 2016; Unterluggauer et al., 2016). In the central Apennines, a general increase of chamaephytes, drought-tolerant species (Petriccione, 2005) and graminoids (Stanisci et al., 2016) was recorded too. Furthermore, graminoids demonstrated better growth performance under a warm and dry climate, which is most likely related to their strategy to allocate resources to belowground parts (Wellstein et al. 2017).

Our results, based on plant cover data recorded in permanent plots, highlight that the observed changes in species composition and structure in alpine and subalpine ridge habitats of the Alps and Apennines are causing a greening trend. Our results are largely consistent with those determined through remote sensing studies (Carlson et al. 2017), which identified on-going greening trends in over half (67%) of the above treeline habitats in the French Alps.

As vegetation cover may be considered a proxy of standing biomass (Fry et al. 2013), its increase affects ecosystem productivity and services in mountain landscapes.

Indeed, climate warming effects, changes in rainfall seasonality and water availability have been proven to be important for ecosystem productivity (Rammig et al., 2010). This issue has been investigated at the Forests of the Apennines (LTER_EU_IT_003 site; Tab. 1) research site where changes in aboveground net primary productivity (ANPP) in response to a shift in the precipitation regime have been detected (Chelli et al., 2016).

Ferretti et al. (2014), in a study including LTER forest sites in the Alps and Apennines, demonstrated that an increase in nitrogen deposition had a positive effect on tree growth (measured as basal area increment) and on aboveground net primary productivity (ANPP), thus promoting carbon sequestration. However, reduction in rainfall can override such positive effects (Chelli et al. 2017). Long-term studies performed at a beech forest in the Apennines (LTER_EU_IT_031; table 1) showed that carbon (C) sequestration depends on both water availability (precipitation) and air temperature. Lower C sequestration in the beech forest was detected in the years characterized by below-average summer precipitation especially when there was also warmer temperatures (Scarascia and Matteucci, 2014; Mazzenga, 2017). Furthermore, in these beech forests, a significant increase in the growing season length and a general increase in the annual net C sequestration were detected from remotely sensed data during 2000 – 2015 (Mazzenga, 2017).

### 3.3 Long-term changes in water chemistry

The studied mountain lakes (Tab. S1) are representative of varying levels of solute content and buffer capacities, from very diluted water (LPS, LPI: conductivity < 10 µS cm$^{-1}$ at 20°C; alkalinity 20-50 µeq L$^{-1}$) to moderately diluted waters (GKS, SCU: conductivity 25-30 µS cm$^{-1}$; alkalinity 100-150 µeq L$^{-1}$) or highly mineralized water (LBS: 60 µS cm$^{-1}$; alkalinity 400-500 µeq L$^{-1}$). All of the lakes are oligotrophic or ultraoligotrophic systems (total phosphorus < 10 µg P L$^{-1}$).

All the lakes showed an increase in alkalinity values (Fig. 6). These trends were highly significant ($p<0.001$) according to the Mann-Kendall test. Slopes varied from 0.8-0.9 µeq L$^{-1}$ y$^{-1}$ (LPI, LPS, GKS) to 3.9 µeq L$^{-1}$ y$^{-1}$ (LBS). In the LPS and LPI, the alkalinity trend was mainly a sign of acidification recovery in response to a decrease in acid deposition after a period of acidification in the 1980s (Rogora et al., 2013). Conductivity also increased over time in the GKS and LBS ($p<0.001$; slopes 0.16 and 0.56 µS cm$^{-1}$ y$^{-1}$, respectively), whereas it slightly decreased in the LPI and LPS. Lake Scuro showed high interannual variability in both conductivity and alkalinity, with a tendency towards increasing values for the latter variable (from 100-120 to 200 µeq L$^{-1}$).

Highly significant negative trends in $SO_4$ concentrations were observed in the LPI and LPS (-6.5 µeq L$^{-1}$ y$^{-1}$), due to a sharp decrease in $SO_4$ deposition that occurred throughout Europe over the last three decades (Rogora et al., 2006). In contrast, $SO_4$ increased significantly in the GKS and LBS (slopes of 3.7 and 6.0 µeq L$^{-1}$ y$^{-1}$, respectively). These trends may be ascribed to an enhanced release of sulphate from the rocks and soils in the catchments of those two lakes. Both lakes also showed positive trends in base cation (calcium and magnesium) concentrations. This increase in the content of major ions in lake water (sulphate, bicarbonate, base cations) has been reported elsewhere in the Alps (Sommaruga-Wögrath et al., 1997; Rogora et al., 2013; Thies et al., 2013; Ilyashuk et al., 2014) and in other remote regions (Williams et al., 2006; Kokelj et al., 2009; Preston et al., 2016; Salerno et al., 2016) where it was ascribed to climate drivers, including a decrease in the amount and extent of snow cover, a shift in ice-off dates, glacier retreat, and permafrost thawing. Low rates of runoff in the summer of dry and warm years may also contribute to concentrated solutes in the runoff water and in the lake (Preston et al., 2016).

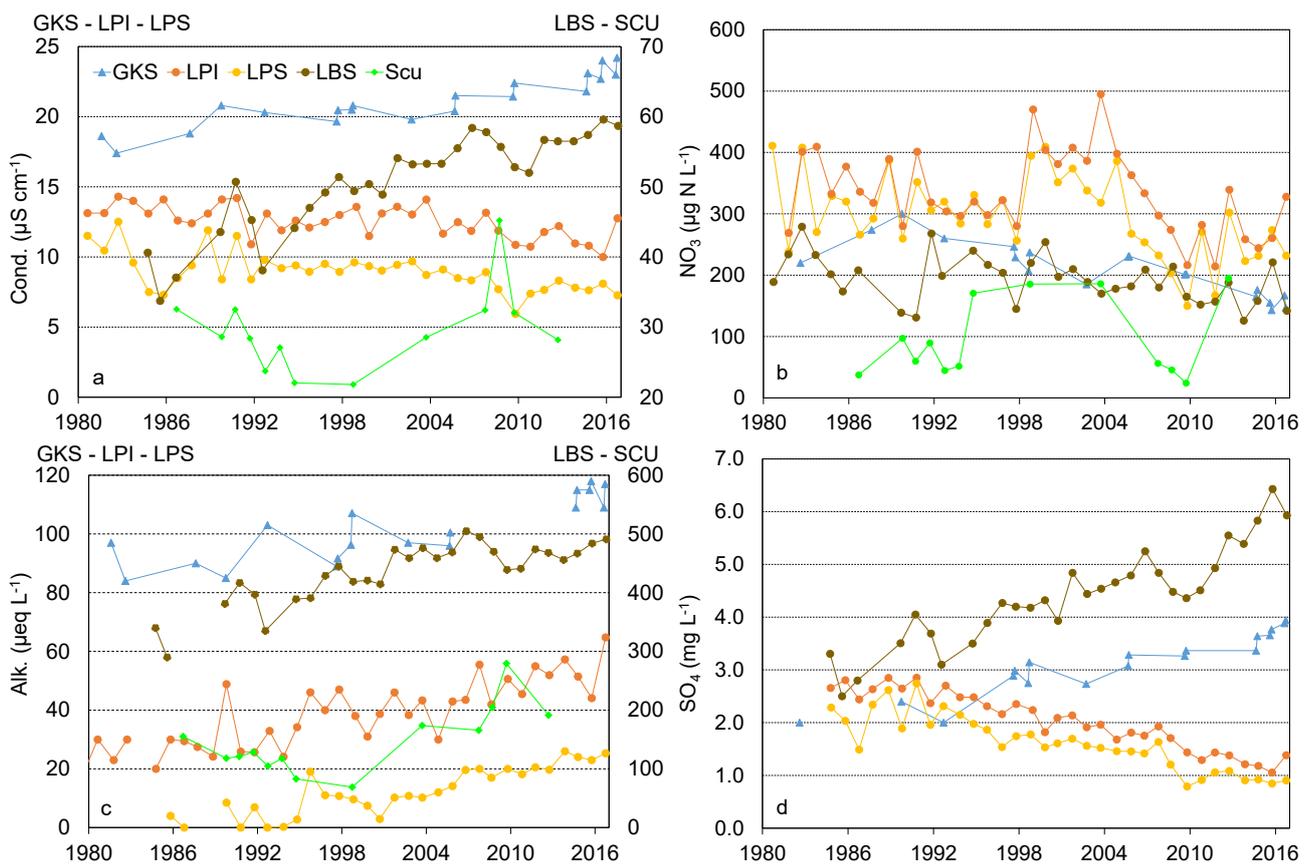

Fig. 6 – Time series of selected water chemical variables at LTER lake sites in the Italian and Austrian Alps and in the Northern Apennines, Italy: a) conductivity at 20°C; b) nitrate; c) alkalinity; d) sulphate. For the lake acronyms, see Tab. S3.

The study sites in the Alps showed a common trend in decreasing nitrate concentrations over the past decade (Fig. 6). The trend was significant at the GKS, LPS (p<0.01), LPI and LBS (p<0.05) sites. The trend slopes varied between -2.0/-2.5 µg L$^{-1}$ y$^{-1}$ in the LPI and LBS and -2.8 µg L$^{-1}$ y$^{-1}$ in the GKS and LPS. Lakes in the Apennines showed a high interannual variability (e.g., NO$_3$ in Lake Scuro varied from 20-30 to 200 µg N L$^{-1}$), without any evidence of a trend.

The negative temporal trends observed for NO$_3$ concentrations could be related to an increase in primary productivity in the lakes in response to climate warming, promoting nitrogen uptake (Sommaruga-Wögrath et al., 1997). Nitrogen uptake also occurs in lake catchments and is regulated by the extent of soil and vegetation (Marchetto et al., 1995). No specific studies have been performed at the lake sites to assess potential changes in plant cover in the catchments. However, it may be hypothesized that the increase in plant cover observed during the last two decades at the LTER vegetation sites (see paragraph 3.2) has taken place in the lake areas too, contributing to the increase in N retention and decrease in NO$_3$ export to the surface waters. Furthermore, changes in depth and extent of snow cover have been shown to affect soil nutrient dynamics (see paragraph 3.1), thereby regulating N release to the water compartment. In general, soil, vegetation and water are strongly interconnected with each other in these high-altitude environments (Magnani et al., 2017); the observed changes in lake water NO$_3$ are probably the result of several interacting processes. Besides climate, a decrease in N input from the atmosphere is likely to have played a role in the observed trend in NO$_3$, especially for acid-sensitive lakes in the Alps; atmospheric deposition of inorganic nitrogen has recently decreased as an effect of decreasing emissions of N compounds, mainly in the oxidized form (Waldner et al., 2014; Rogora et al., 2016). The effects of decreasing N deposition on NO$_3$ levels in rivers and lakes have been widespread, with several monitoring sites in Europe showing a significant negative trend in NO$_3$ concentrations (Garmo et al., 2014).

### 3.4 Biological response in lakes and rivers

*3.4.1 Large-scale patterns*

The effect of a large-scale climatic pattern (EA: East Atlantic pattern climatic index) on plankton phenology was tested in two Apennine lakes (LTER_EU_IT_046 and LTER_EU_IT_047; Tab. S3).

In Lake Santo, we found a positive correlation between the spring EA values and June water temperatures (Spearman's rho=0.66, $p < 0.05$; not shown in tables), while in Lake Scuro, the spring EA was positively correlated with May water temperatures (Spearman's rho=0.67, $p < 0.05$; not shown in tables). In Lake Santo, we observed significant negative correlations between the spring EA values and the center of gravity calculated for chlorophyll-*a* at the surface, intermediate, and bottom layers, respectively. We also found significant negative correlations between the spring EA values and the date of the first appearance of several of the most abundant zooplankton taxa, including the dominant microcrustaceans (*Daphnia longispina*, *Bosmina longirostris*, and *Eudiaptomus intermedius*) (Tab. 2).

We did not find significant relationships between the proxies of phytoplankton and zooplankton phenology (chlorophyll-*a* center of gravity, date of peak of chlorophyll-*a* concentration, dates of first seasonal appearance of zooplankton taxa) in Lake Santo. In Lake Scuro, the dates of the first seasonal appearance of several zooplankton taxa were positively related to the corresponding dates of peak chlorophyll concentrations (Tab. 2), while both the phytoplankton and zooplankton phenological proxies did not show any relationship with the EA values.

Our results for Lake Santo suggest that large-scale climate variations may be associated with changes in lake plankton phenology, likely because of changes in local climate (e.g., air temperature) and lake thermal dynamics. Specifically, in Lake Santo, we observed that years with positive spring EA values were characterized by higher spring air temperatures, warmer June water temperatures, and earlier seasonal development of both phytoplankton (represented by chlorophyll-a) and key zooplankton taxa. The lack of similar climatic signatures on plankton dynamics in Lake Scuro is most likely due to the relatively small size and substantially reduced thermal inertia of this lake, resulting in a markedly higher sensitivity of Lake Scuro to local meteorological variability and thereby masking the potential impacts of large-scale climatic patterns when compared to Lake Santo. On the other hand, we found significant correlations between the phytoplankton and zooplankton proxies in Lake Scuro, indicating a stronger influence of interspecific interactions in this lake.

Tab. 2 - Spearman's correlation values between: a) spring EA values and plankton phenology metrics for Lake Santo; b) phyto- and zooplankton phenology metrics for Lake Scuro. CG: chlorophyll-a center of gravity calculated for the surface (sur), intermediate (int) and bottom (bot) layers of the water column; ChlaMax: date of peak chlorophyll-a concentration; C1, 2, 3: 1st, 2nd, and 3rd copepodite stages; ns: not significant.

|  | EAspr (SAN) | ChlaMax (SCU) |
|---|---|---|
| **Phytoplankton** | | |
| CGsur | -0.68 | - |
| CGint | -0.70 | - |
| CGbot | -0.75 | - |
| **Zooplankton (date of 1st appearance)** | | |
| *Eudiaptomus intermedius ovigerous* F | -0.79 | ns |
| *Eudiaptomus intermedius nauplii* | -0.80 | ns |
| *Eudiaptomus intermedius* C1 | ns | 0.77 |
| *Eudiaptomus intermedius* C2 | ns | 0.68 |
| *Eudiaptomus intermedius* C3 | ns | 0.68 |
| *Cyclopoid copepodites* | ns | 0.66 |
| *Mesocyclops leuckarti* | ns | 0.69 |
| *Conochilus gr. unicornis-hippocrepis* | ns | 0.66 |
| *Daphnia longispina* | -0.80 | ns |
| *Bosmina longirostris* | -0.76 | ns |
| *Ascomorpha ecaudis* | -0.69 | ns |
| *Kellicottia longispina* | -0.69 | ns |
| *Keratella quadrata* | -0.69 | ns |
| *Synchaeta gr. stylata-pectinata* | -0.69 | ns |
| **Zooplankton (date of peak)** | | |
| *Eudiaptomus intermedius nauplii* | ns | 0.87 |
| *Eudiaptomus intermedius* C2 | ns | 0.62 |
| *Keratella cochlearis* | ns | 0.57 |
| Cyclopoid nauplii | ns | 0.69 |
| *Pleurata sp.* | ns | 0.17 |

Overall, these results indicate that large-scale climate indices can be useful indicators of climate variation at a local scale. The EA, in particular, was confirmed as a relevant index for the Mediterranean area (Salmaso et al., 2012). However, the response of lakes to large-scale climatic patterns is largely dependent on the thermal structure and mixing regime of lakes (Gerten and Adrian, 2001).

*3.4.2 Local patterns*

The effects of local patterns, mainly abiotic parameters, on benthic communities were investigated at the LTER site Matsch/Mazia Valley (LTER_EU_IT_100). Long-term data showed clear seasonal distribution patterns; the increased discharge at snowmelt during June and July led to a sharp decrease of faunal density and number of taxa (Fig. 7). A significant negative correlation was found between the monthly discharge and number of individuals (r= -0.73; p<0.001). The presence of a glacier within the drainage basin also played a role, by affecting abiotic parameters (primarily water discharge) over a wide range of time scales, with fundamental implications for the whole river system.

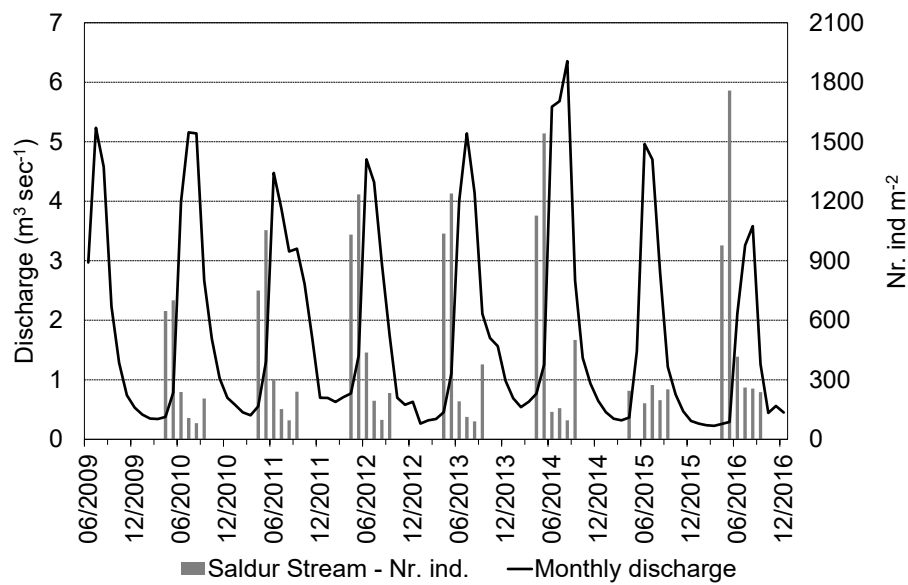

Fig. 7 – Time series of monthly mean discharge and number of macrobenthic individuals at one sampling station located at 2340 m a.s.l. on the Saldur Stream (Matsch/Mazia Valley; LTER_EU_IT_100).

These results demonstrate that, in addition to large-scale patterns, minor changes in environmental conditions can also have significant consequences on biological communities in mountain lakes and streams (Wallace and Webster, 1996). In particular, the snow melting process significantly modifies the biological communities of stream invertebrates in the Saldur stream (Fig. 7). Benthic community structure in mountain streams is a result of complex environmental interactions (Milner et al. 2001; Zemp at al., 2009; Lencioni and Spitale, 2015; Niedrist and Füreder, 2017). Therefore, the understanding of hydro-ecological relationships is essential for the development of effective conservation strategies for alpine rivers. Long-term observations on benthic invertebrate communities may enable assessments of the potential impacts of global change on stream ecosystems (Jourdan et al., 2018). In particular, the community composition of the numerous small alpine valleys, which are often not investigated from the faunistic point of view, may be an important proxy for environmental changes, including climate change.

Besides the long-term effects of climate change, high-elevation ecosystems may also be affected by extreme climatic events such as heat waves, droughts, heavy rainfall and floods (Jones, 2013). Studies at the survey lakes in the Western Alps, Italy, belonging to the parent site LTER_EU_IT_009 Mountain Lakes, showed that climatic factors, particularly air temperature and SCD, interact with atmospheric deposition and determine short-term changes in lake water e.g., heavy rainfall or snowy winters caused a temporary decrease in the alkalinity pool in the lakes by dilution and a simultaneous pulse of $NO_3$ to the lake, with an overall acidification effect (Rogora et al., 2013). The impact of extreme climatic events on ecological processes was also addressed at the LTER site LTER_EU_IT_047 Lake Scuro Parmense; Bertani et al. (2016) observed a shift from an unvegetated to a macrophyte-dominated regime as a result of the 2003 heat wave. Some of the observed changes in the lake food web persisted after 2003, suggesting that abrupt and long-lasting ecosystem-level reorganizations may occur in small mountain lakes as an effect of extreme events.

**Conclusions**

The long-term ecological analysis we performed, based on data obtained from permanent plots, provided evidence that mountain ecosystems in the Alps and Apennines, both terrestrial and freshwater, show varying levels of effects in response to climate change.

The results of our analyses and the review of the results gathered at the study sites from previous and on-going studies highlight that climate change effects are mainly indirect and result from multiple, interacting processes. To assess these changes, there is a need for strong partnerships in mountain ecosystem observation and research and for multidisciplinary approaches, encompassing the distinction between different types of ecosystems (Mirtl et al., 2018).

The observed long-term ecological changes include the increase in vegetation cover and in soil microbial biomass in alpine and subalpine summits, and the increase in C uptake in mountain forests. The interannual variability in snow cover duration plays a relevant role in nutrient cycles, both in soils and in surface waters, and snow cover change, when coupled with climate-related vegetation phenology, was also proven to affect animal population dynamics, namely, some glacial follower species, such as the alpine ibex. Snow- and ice-melting processes also affect biological communities of glacier-fed streams by interacting with abiotic parameters such as water discharge and turbidity.

In addition to long-term changes, short-term episodes or extreme events also proved to be relevant for mountain ecosystems, causing, for instance, a regime shift in response to heat waves, pulses of nutrient or chemicals to lake water at snowmelt and sudden changes in the nutrient dynamics in soils.

The joint analyses we provided demonstrate that long-term research is essential to understanding mountain ecosystem complexity and dynamic. The results also highlighted the great potential for further scientific advances that rely on international collaboration and integration. From this perspective, the LTER is an ideal network for improving our knowledge on sensitive ecosystems such as mountain soils, vegetation and freshwater lakes and streams.

Our effort of combining a huge amount of data gathered from different ecosystem types also demonstrates the limits of such an approach; there is a strong need for adopting co-located monitoring site networks (Haase et al. 2018) to improve our ability to obtain sound results from cross-site analysis. Moreover, a useful tool would consists of the adoption of site and dataset registries, providing access to site metadata and information on existing collaborative networks and research platforms.

Nevertheless, there is a need for further studies, in particular, short-term analyses with fine spatial and temporal resolutions to improve our understanding of the response to extreme events and an effort to increase comparability and standardize protocols across networks to clarify local patterns from global patterns.

The outcomes of this paper demonstrate that LTER mountain sites would gain additional value from the development and improvement of joint networks, indicators, and methodologies. This approach would take advantage of mountain ecosystems as early warning indicators in monitoring frameworks.

**Acknowledgments**

This study was supported by the following projects: Project of Interest NextData trough the Special Project Data-LTER-Mountain (Harmonisation and standards for existing and newly collected Data and MetaData on LTER sites in Italian Mountain ecosystems); H2020 EcoPotential (grant agreement No. 641762), eLTER H2020 (GA n. 654359) and Advance_eLTER (GA n739558); project MEDIALPS - Disentangling anthropogenic drivers of climate change impacts on alpine plant species: Alps vs. Mediterranean mountains, funded by the Austrian Academy of Sciences; LTSER Austria, GLEON, and the University of Innsbruck. We thank three anonymous reviewers for providing helpful comments and suggestions on the manuscript.

**Conflict of interest**

The authors certify that there is no actual or potential conflict of interest in relation to this article.